\documentclass[11pt]{article}

\usepackage{acl}

\usepackage{times}
\usepackage{latexsym}
\usepackage[T1]{fontenc}
\usepackage[utf8]{inputenc}
\usepackage{microtype}
\usepackage{inconsolata}
\usepackage{graphicx}
\usepackage{url}

\title{Diagnosable ColBERT: Debugging Late-Interaction Retrieval Models \\ Using a Learned Latent Space as Reference}

\author{Remy Fran\c{c}ois \\
Parallia AI \\
francois.remy@parallia.eu}

\begin{document}
\maketitle

\begin{abstract}
Reliable biomedical and clinical retrieval requires more than strong ranking performance: it requires a practical way to find systematic model failures and curate the training evidence needed to correct them.
Late-interaction models such as ColBERT \citep{khattab2020colbert} provide a first solution thanks to the interpretable token-level interaction scores they expose between document and query tokens.
Yet this interpretability is shallow: it explains a particular document--query pairwise score, but does not reveal whether the model has learned a clinical concept in a stable, reusable, and context-sensitive way across diverse expressions.
As a result, these scores provide limited support for diagnosing misunderstandings, identifying irreasonably distant biomedical concepts, or deciding what additional data or feedback is needed to address this.
In this short position paper, we propose Diagnosable ColBERT, a framework that aligns ColBERT token embeddings to a reference latent space grounded in clinical knowledge and expert-provided conceptual similarity constraints.
This alignment turns document encodings into inspectable evidence of what the model appears to understand, enabling more direct error diagnosis and more principled data curation without relying on large batteries of diagnostic queries.
\end{abstract}

\section{Introduction}


\begin{figure*}[t]
\centering
\includegraphics[width=\textwidth]{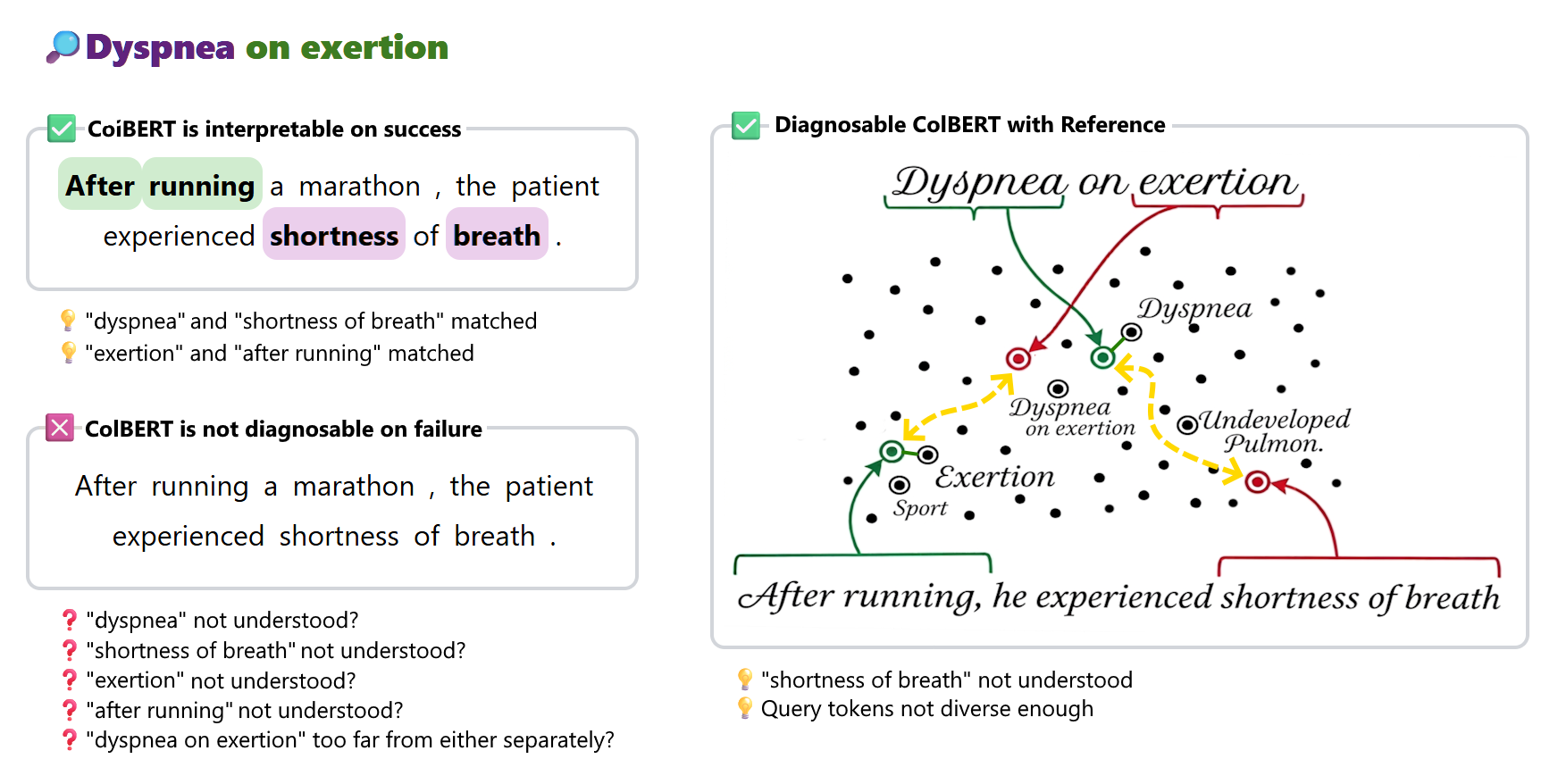}
\caption{Standard ColBERT-style late interaction produces interpretable query--document scores, but provides little guidance when no meaningful match is found for a token, as illustrated on the left. By contrast, augmenting the retriever with a reference latent space, as on the right, allows testers of Diagnosable ColBERT models to diagnose failures more quickly and more actionably by separating latent-space semantics from in-context understanding.}
\label{fig:diagnosable-colbert-overview}
\end{figure*}

Trustworthy artificial intelligence in high-stakes domains increasingly requires transparent reporting of model capabilities, limitations, and failure modes \citep{who2021ethics,eu2017mdr,abridge2024evaluation}.
Finding systematic model failures is a central requirement for deploying retrieval systems in clinical and biomedical settings.
When a retrieval model confuses nearby concepts, overgeneralizes across contexts, or fails to encode clinically salient negation, the consequence is not merely a lower ranking metric, but a weaker basis for downstream decision making and a poorer signal for curating corrective training data \citep{harkema2009context,lee2010encoding,wang2020medsts}.
For this reason, debugging should ideally reveal not only that a model failed on a particular document--query pair, but also \emph{what} the model appears to understand, \emph{where} that understanding breaks down, and \emph{which} additional examples are needed to improve it.

Late-interaction models in the style of ColBERT are appealing in this regard because they expose fine-grained interaction scores between query and document tokens \citep{khattab2020colbert}.
However, this evidence remains retrospective and local: it helps explain why a given query matched a given document, but not whether the model has learned clinically meaningful distinctions robustly across paraphrases, contexts, and compositional variants \citep{huang-baldwin-2023-robustness,kang-etal-2025-trial}.
Explaining a match is not the same as gauging and diagnosing misunderstandings.

In clinical retrieval, the most consequential failures often depend on context-sensitive factors such as negation, temporality, uncertainty, experiencer, historical status, and the distinction between mentioning a test and asserting a condition.
These are precisely the kinds of distinctions that may be weakly expressed in local match scores while remaining decisive for retrieval quality.

We therefore argue that checking model understanding requires interpretation at multiple levels of contextualization.
Many biomedical meanings are only partially specified at the token level and become clearer as progressively more context is introduced \citep{lee2010encoding}.
For example, a model that appears reasonable on \emph{cat} may still fail on \emph{cat scratch disease}. Moreover, paragraph-level context can further influence whether a mention is current, historical, speculative, negated, or even concerns another person \citep{harkema2009context,wang2020medsts}.
Interpretability must therefore extend beyond isolated token matches toward a graded view of understanding across term, sentence, and paragraph-level alignment.
Accordingly, we treat diagnostic alignment as multi-factorial: it should reveal term-level concept identity, phrase-level composition, context-level qualifiers, and broader discourse elements that alter clinical interpretation.

This broader view of interpretation also motivates the use of a structured latent space rather than a fixed inventory of labels.
If the embedding space is shaped to respect concept-level and sentence-level similarity constraints, then it can represent not only previously named clinical concepts, but also contextually enriched and partially novel combinations of them \citep{campbell2014semantic}.
Prior work also suggests that such spaces are feasible to construct in practice, whether by grounding biomedical representations in ontological definitions and relations as in BioLORD \citep{remy2024biolord}, by aligning dense representations to structured label spaces \citep{decorte2025efficient}, or by grounding annotated mentions in a dynamic model-generated concept space when the concept inventory is large and evolving \citep{stepanov2026glinerbiencoder}.

\newpage
In this publication, we therefore propose Diagnosable ColBERT, a framework that leverages such a reference latent space learned from expert-specified similarity constraints over clinical concepts, and aligns late-interaction token embeddings to it via pre-projection adapters.
The central proposal of this paper is therefore not a new retrieval scoring rule, but a new diagnostic lens: late-interaction representations should be interpreted against a clinically grounded reference space rather than only through pairwise match scores.
The goal is not merely to explain a retrieval score after the fact, but to make document encodings themselves interpretable as evidence about what the model appears to know.
This creates a path toward diagnosing misunderstandings directly from encoded documents, identifying confusable concepts, and curating training data in a more principled way.
More generally, we advocate a shift from retrospective score inspection to concept-grounded diagnosis of model understanding.
Such representations could support both direct human inspection and more automated auditing workflows.

\newpage
\section{Diagnostic Framework}


Diagnosable ColBERT is organized around a pre-existing reference latent space that serves as a domain-specific diagnostic scaffold. Similarly to BioLORD \citep{remy2024biolord}, this latent space needs to accommodate both concept names, clinical sentences, and paragraphs.
The purpose of this space is to make contextual token representations clinically legible: not only in terms of term-level concept identity, but also in terms of local composition and context-level qualifiers such as negation, temporality, uncertainty, or experiencer.
Instead of only asking which query token matched which document token, we ask what clinically relevant factors a contextualized representation appears to encode and how they arrange geometrically.

In this view, alignment means mapping late-interaction token representations into a space where these factors can be inspected more directly.
The retrieval representation should remain tied to this diagnosed representation, but need not be identical to it.
A natural design is for retrieval embeddings to be learned as a lower-dimensional downprojection of the diagnosed representation, so that retrieval can reweight clinically relevant factors for ranking efficiency without discarding the richer structure needed for diagnosis.
This keeps the framework centered on a simple idea: retrieval and diagnosis should stay coupled, but diagnosis should not be reduced to whatever geometry happens to be most convenient for fast search.

The specific architecture used to realize this idea is secondary to the paper's main claim.
One can implement the diagnostic view with lightweight contextual adapters and task-specific projection heads; our essential contribution is the diagnostic framing itself: a clinically grounded reference geometry for inspecting what a late-interaction retriever appears to perceive from a context.

\begin{figure}[t]
\centering
\includegraphics[width=\linewidth]{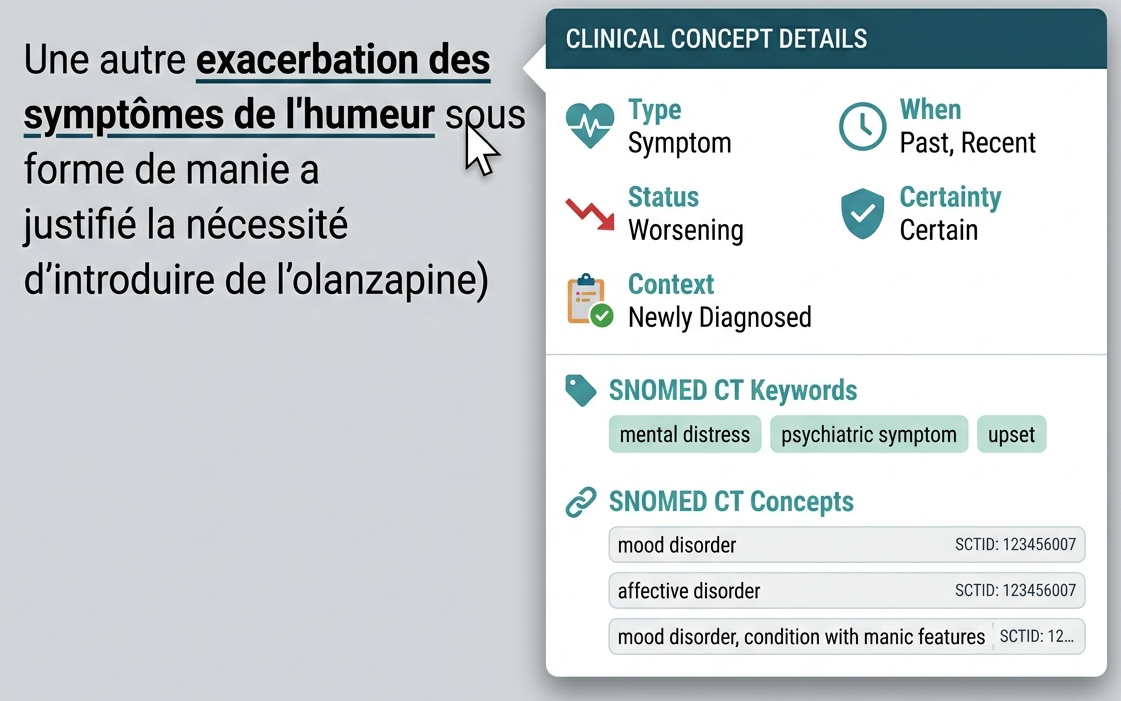}
\caption{Proposed debugging interface for Diagnosable ColBERT. The interface exposes token-level query and document representations together with their placement in the reference latent space and the clinically meaningful concepts or qualifiers associated with nearby regions. This gives testers a practical way to distinguish weak interaction scores from deeper failures of concept grounding, abbreviation handling, or contextual interpretation, and to turn a retrieval miss into an actionable diagnosis.}
\label{fig:diagnosable-colbert-interface}
\end{figure}

\section{Practical Examples}

A central limitation of standard ColBERT interpretability is that it is fundamentally relational: it explains why a query matched a document, but offers limited leverage when the goal is to determine why a relevant match failed to occur (see Figure \ref{fig:diagnosable-colbert-overview}).
Diagnosable ColBERT is intended to address precisely this gap.
Rather than treating retrieval failure as a single undifferentiated event, it seeks to identify which part of the representation pipeline failed to preserve the clinically relevant meaning.

Consider a case report retrieval system in which a tester issues the query \emph{bartonellosis}.
A relevant report is missed because the report mentions only \emph{CSD}, the abbreviation for \emph{cat scratch disease}.
Standard query--document inspection establishes that the interaction between \emph{bartonellosis} and \emph{CSD} is too weak.
But this finding alone is diagnostically incomplete.
It does not tell us whether the query representation failed to capture the target disease family, whether the document representation failed to interpret the abbreviation correctly, or whether both encodings are deficient.

Diagnosable ColBERT resolves this ambiguity by grounding both sides in a reference latent space.
The tester can inspect whether \emph{bartonellosis} is already positioned near the relevant disease concept and, separately, whether \emph{CSD} is mapped into that same region.
If \emph{bartonellosis} is correctly grounded but \emph{CSD} is not, then the root cause is not weak interaction per se, but document-side abbreviation understanding.
If both are misplaced, the problem lies deeper, at the level of concept representation.
This kind of distinction is difficult to obtain from pairwise interaction scores alone, yet it directly guides remediation.

When the problem lies in abbreviation grounding, the natural intervention is to curate examples that explicitly connect abbreviated, expanded, and synonymous forms. 
When the problem lies in broader concept placement, the remedy instead concerns the shaping of the underlying semantic space. 

The same diagnostic logic extends beyond query-conditioned retrieval.
Some failures can be identified directly from the document encoding itself.
Consider the utterance: \emph{Are you allergic to any of the following drugs? --- Yes, to ranitidine}.
Here the relevant question is whether the token representation of \emph{ranitidine} reflects not only the medication identity, but also the allergic relation introduced by context.
If the encoding remains close to the drug concept while failing to capture the neighboring context of allergy, adverse reaction, or intolerance, then the model has encoded the entity but not the clinically decisive context.

\vspace{0.225cm}
This is a query-free diagnostic signal: the tester can identify the failure before retrieval is even attempted.
Again, the diagnosis suggests a targeted intervention, namely the addition of training examples in which drug mentions appear indirectly in contexts related to allergy or adverse-reactions, and ensure that this context is getting properly picked up by the encoder model.

\vspace{0.25cm}
\section{Real-World Usage}
In our experience, debugging user interfaces for Diagnosable ColBERT models\footnote{\mbox{Public demos: \url{http://demo26.parallia.eu/} and \url{http://text2json.parallia.eu}.}} have already proven useful as a practical way to inspect token-level phenomena and surface failures that would have been difficult to localize from ranking metrics alone.
What made the interface valuable was that it helped spot remaining representational weaknesses that actually mattered in practice, including gaps in conceptual grounding, abbreviation and brand name handling, and preservation of clinical context.

\vspace{0.225cm}
Subjectively, this changed the development process.
Instead of treating misretrievals as abstract evaluation outcomes, we were often able to trace issues back to recurring representational problems and then respond with more targeted curation and model changes.
These observations informed the progression from ClinicalEncoder25 to ClinicalEncoder26AM and beyond, even if this workflow remains work in progress and should not be mistaken for a complete evaluation methodology.

\newpage
\section{Limitations and Scope}

Diagnosable ColBERT is best understood as a diagnostic scaffold, not as a complete semantics of biomedical meaning.
Its purpose is to make clinically relevant aspects of the model's internal organization more legible to a tester, not to claim that every retrieval-relevant distinction can be exhaustively captured in a single reference geometry.
Any such space reflects design choices about which concepts, relations, and contextual qualifiers are made salient, and its usefulness depends on being well shaped by domain knowledge, expert constraints, and clinically meaningful similarity structure.
The proposal therefore improves diagnosability without by itself guaranteeing retrieval quality: diagnostic alignment complements retrieval evaluation by making failure modes more interpretable and actionable, but useful systems will still need strong ranking behavior as well.

\section{Conclusion}

In this paper, we have argued that debugging late-interaction retrievers requires more than inspecting token-level query--document scores after a failure has already occurred.
Our central claim is that retrieval models become substantially more diagnosable when their token representations are aligned to a clinically grounded reference latent space that makes concept grounding, contextual qualifiers, and representational gaps directly inspectable.

Diagnosable ColBERT is a concrete step in this direction.
Rather than treating retrieval as a black box whose failures must be inferred indirectly from weak matches and downstream metrics, it provides a practical framework for localizing why a miss occurred and for inspecting difficult document segments even before a query is issued.

Much remains to be refined, especially in constructing strong reference spaces and turning diagnoses into robust improvement workflows. Even so, we view the direction as already promising in practice: clinically grounded diagnostic representations complement ranking evaluation by making retrieval behavior more legible and actionable.



\newpage
\bibliography{custom}

@misc{who2021ethics,
  author = {{World Health Organization}},
  title = {Ethics and Governance of Artificial Intelligence for Health: {WHO} Guidance},
  year = {2021},
  month = jun,
  publisher = {World Health Organization},
  url = {https://www.who.int/publications/i/item/9789240029200},
  note = {Published June 28, 2021}
}

@misc{eu2017mdr,
  author = {{European Parliament and Council of the European Union}},
  title = {Regulation ({EU}) 2017/745 of the European Parliament and of the Council of 5 April 2017 on Medical Devices, Amending Directive 2001/83/{EC}, Regulation ({EC}) No 178/2002 and Regulation ({EC}) No 1223/2009 and Repealing Council Directives 90/385/{EEC} and 93/42/{EEC}},
  year = {2017},
  month = may,
  howpublished = {Official Journal of the European Union, L 117, pp. 1--175},
  url = {https://eur-lex.europa.eu/legal-content/ENG/TXT/?uri=CELEX:32017R0745},
  note = {Adopted April 5, 2017; published May 5, 2017}
}

@misc{abridge2024evaluation,
  author = {Michael Oberst and Davis Liang and Zachary C. Lipton},
  title = {Pioneering the Science of {AI} Evaluation},
  year = {2024},
  month = sep,
  publisher = {Abridge},
  url = {https://www.abridge.com/ai/science-ai-evaluation},
  note = {Whitepaper, published September 19, 2024; updated August 7, 2025}
}

@inproceedings{khattab2020colbert,
  author = {Omar Khattab and Matei Zaharia},
  title = {{ColBERT}: Efficient and Effective Passage Search via Contextualized Late Interaction over {BERT}},
  booktitle = {Proceedings of the 43rd International {ACM} {SIGIR} Conference on Research and Development in Information Retrieval},
  pages = {39--48},
  year = {2020},
  doi = {10.1145/3397271.3401075}
}

@article{harkema2009context,
  author = {Hendrik Harkema and John N. Dowling and Tyler Thornblade and Wendy W. Chapman},
  title = {{ConText}: An Algorithm for Determining Negation, Experiencer, and Temporal Status from Clinical Reports},
  journal = {Journal of Biomedical Informatics},
  volume = {42},
  number = {5},
  pages = {839--851},
  year = {2009},
  doi = {10.1016/j.jbi.2009.05.002}
}

@article{lee2010encoding,
  author = {Dennis H. Lee and Francis Y. Lau and Hue Quan},
  title = {A Method for Encoding Clinical Datasets with {SNOMED CT}},
  journal = {BMC Medical Informatics and Decision Making},
  volume = {10},
  pages = {53},
  year = {2010},
  doi = {10.1186/1472-6947-10-53}
}

@article{wang2020medsts,
  author = {Yanshan Wang and Naveed Afzal and Sunyang Fu and Liwei Wang and Feichen Shen and Majid Rastegar-Mojarad and Hongfang Liu},
  title = {{MedSTS}: A Resource for Clinical Semantic Textual Similarity},
  journal = {Language Resources and Evaluation},
  volume = {54},
  number = {1},
  pages = {57--72},
  year = {2020},
  doi = {10.1007/s10579-018-9431-1}
}

@article{remy2024biolord,
  author = {Fran{\c{c}}ois Remy and Kris Demuynck and Thomas Demeester},
  title = {{BioLORD}-2023: Semantic Textual Representations Fusing Large Language Models and Clinical Knowledge Graph Insights},
  journal = {Journal of the American Medical Informatics Association},
  volume = {31},
  number = {9},
  pages = {1844--1855},
  year = {2024},
  doi = {10.1093/jamia/ocae029}
}

@article{decorte2025efficient,
  author = {Jens-Joris Decorte and Jeroen Van Hautte and Chris Develder and Thomas Demeester},
  title = {Efficient Text Encoders for Labor Market Analysis},
  journal = {IEEE Access},
  volume = {13},
  pages = {133596--133608},
  year = {2025},
  doi = {10.1109/ACCESS.2025.3589147}
}

@misc{stepanov2026glinerbiencoder,
  author = {Ihor Stepanov and Mykhailo Shtopko and Dmytro Vodianytskyi and Oleksandr Lukashov},
  title = {The Million-Label {NER}: Breaking Scale Barriers with {GLiNER} bi-encoder},
  year = {2026},
  eprint = {2602.18487},
  archivePrefix = {arXiv},
  primaryClass = {cs.CL},
  url = {https://arxiv.org/abs/2602.18487},
  note = {Introduces {GLiNKER}, a modular framework for large-scale entity linking}
}

@article{campbell2014semantic,
  author = {Walter S. Campbell and James R. Campbell and William W. West and James C. McClay and Steven H. Hinrichs},
  title = {Semantic Analysis of {SNOMED CT} for a Post-Coordinated Database of Histopathology Findings},
  journal = {Journal of the American Medical Informatics Association},
  volume = {21},
  number = {5},
  pages = {885--892},
  year = {2014},
  doi = {10.1136/amiajnl-2013-002456},
  pmid = {24833774}
}

@inproceedings{huang-baldwin-2023-robustness,
  title = {Robustness Tests for Automatic Machine Translation Metrics with Adversarial Attacks},
  author = {Huang, Yiran and Baldwin, Timothy},
  booktitle = {Findings of the Association for Computational Linguistics: EMNLP 2023},
  pages = {4649--4675},
  year = {2023},
  doi = {10.18653/v1/2023.findings-emnlp.340},
  url = {https://aclanthology.org/2023.findings-emnlp.340/}
}

@inproceedings{kang-etal-2025-trial,
  title = {{TRIAL}: Token Relations and Importance Aware Late-Interaction for Accurate Text Retrieval},
  author = {Kang, Junmo and Ro, Yunhyeok and Heo, Junsie and Seo, Minjoon},
  booktitle = {Proceedings of the 2025 Conference on Empirical Methods in Natural Language Processing},
  pages = {15404--15427},
  year = {2025},
  doi = {10.18653/v1/2025.emnlp-main.854},
  url = {https://aclanthology.org/2025.emnlp-main.854/}
}

\end{document}